\documentclass{article}

\usepackage[english]{babel}

\usepackage{amsmath}
\usepackage{graphicx}
\usepackage[colorlinks=true, allcolors=blue]{hyperref}
\usepackage{amsthm}
\usepackage{multicol}
\usepackage{amsfonts}

\usepackage{tikz}
\usetikzlibrary{automata,arrows,positioning}
\usetikzlibrary{fit, arrows.meta}

\newtheorem{theorem}{Theorem}[section]

\newtheorem{lemma}[theorem]{Lemma}
\newtheorem{definition}{Definition}[section]

\title{Unattainability of Common Knowledge in Asymmetric Games with Imperfect Information} 

\author{Fabian Farestam \\
        ETH Zürich \\
        \href{https://orcid.org/0009-0003-6037-803X}{ORCID: 0009-0003-6037-803X}
        \and 
        Dilian Gurov \\
        KTH Royal Institute of Technology, Stockholm \\
        \href{https://orcid.org/0000-0002-0074-8786}{ORCID: 0000-0002-0074-8786}
        }

\date{\today}

\begin{document}

\maketitle

\begin{abstract}

In this paper, we present a conceptual model game to examine the dynamics of asymmetric interactions in games with imperfect information. The game involves two agents with starkly contrasting capabilities: one agent can take actions but has no information of the state of the game, whereas the other agent has perfect information of the state but cannot act or observe the other agent's actions. This duality manifests an extreme form of asymmetry, and how differing abilities influence the possibility of attaining common knowledge. Using Kripke structures and epistemic logic we demonstrate that, under these conditions, common knowledge of the current game state becomes unattainable. Our findings advance the discussion on the strategic limitations of knowledge in environments where information and action are unevenly distributed.

\end{abstract}


\section{Introduction}

In game theory, the concept of \emph{common knowledge}—where every participant is aware of a piece of information and knows that everyone else is aware of it too—is fundamental for predicting outcomes in strategic interactions. For action coordination, the difference between common knowledge and ``almost" common knowledge can drastically alter outcomes \cite{Rubinstein_1989}, underscoring its critical importance. However, establishing common knowledge can be particularly challenging in asymmetric games with \emph{imperfect information}.

This paper presents a conceptual model game that captures the essence of such challenges through a simple, yet illustrative scenario involving an ambiguous alarm clock interface manipulated by a human, and observed by an AI. The primary contribution of this work is the development of a proof, utilising epistemic logic, that demonstrates the unattainability of common knowledge in certain configurations of the model game.

Scenarios involving multiple agents with limited communication have been analysed extensively. For instance, in the well-known \emph{coordinated attack problem} (also known as the two generals problem), common knowledge cannot be achieved due to unreliable communication channel~\cite{FaginReasoningAboutKnowledge}. Similar games, where achievement of common knowledge is unlikely were investigated in~\cite{Bergmark2022InvestigationStability}; however, the focus of that work was not on analysing whether common knowledge is attainable or not.

The model game presented in the next section focuses on a situation without any form of communication, where two agents can observe different things: one only the actions it takes, whereas the other only the outcomes of these actions, while not taking any actions on its own. This creates a ``distilled" game, for which we formally prove that \emph{common knowledge is unattainable} for certain sequences of actions.
The game is of further interest, since it is the smallest game exhibiting the discussed phenomenon of its kind, namely a \emph{multi-agent game with imperfect information against nature} (MAGIIAN)~\cite{gur-gor-lun-22-aij}; it is smallest in terms of agents, nodes, actions and indistinguishability relations.

The results and conclusions of this study extend beyond theoretical interest, addressing real-world scenarios where similar dynamics may occur. Demonstrating the unattainability of common knowledge in various setups underscores the challenges of achieving sufficient action coordination in systems characterised by limited communication and imperfect information. This insight is particularly vital in designing systems where action coordination is essential, making our methods directly relevant to practical applications. Understanding the attainability of common knowledge is crucial for effectively constructing these systems, providing a foundational strategy for managing complexity in decentralised decision-making environments.


\section{A two-agent game with imperfect information}
\label{game_set_up}

\begin{figure}[t]
    \centering
    \begin{tikzpicture}[->,>=stealth',shorten >=1pt,auto,node distance=5cm,
                        semithick]
      \tikzstyle{every state}=[fill=white,draw=black,text=black]
    
      \node[initial,state] (A)                    {Start};
      \node[state]         (B) [below left of=A]  {On};
      \node[state]         (C) [below right of=A] {Off};
      
      \path (A) edge              node [left] {morning arrives} (B)
                edge              node {morning arrives} (C)
            (B) edge [bend left]   node {reset, toggle} (C)
            (C) edge [bend left]   node {toggle} (B)
                edge [loop right]  node {reset} (C);
    
      \draw [dashed,red,-] (B) -- (C) node[midway, below] {$\sim$ Human};
    \end{tikzpicture}
    \caption{The scenario presented as a game graph.}
    \label{fig:state_transition}
\end{figure}
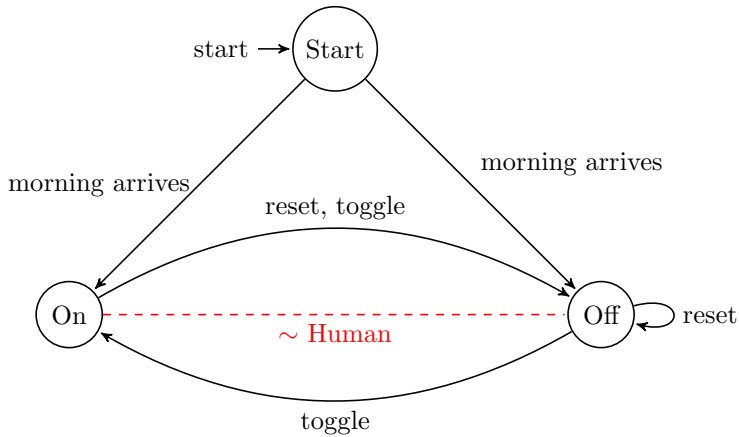

Our scenario, illustrated as a game graph in Figure~\ref{fig:state_transition}, is set up with the following story.
The year is 2052, and everyone's personal AI assistant has even access to the memory of one's alarm clocks.
The alarm clock has two buttons: a ``reset" button, which turns the alarm off, and an ``on/off" button, which toggles the state of the alarm. Due to cost savings, the clock does not indicate if the alarm is on or off. A further complication is that the human can move in their sleep and accidentally press some button and as a result, the human can never be sure if the alarm is on or off in the morning; this uncertainty of the human is modelled by the dashed line in Figure~\ref{fig:state_transition}.

The AI has seen a pattern in the data, namely that the human's mood can be fairly well estimated by knowing if the human knowingly turns the alarm on, turns it off, or just gambles if it is on or off.

An observant human has realised that the AI uses this to estimate their mood, so the human tries to estimate the knowledge of the AI, to hack the behaviour of the AI. This poses a problem since there is a feedback loop in the knowledge states. Thus, the question arises, whether the AI can always truly know the knowledge state of the human.

In this study, we shall prove the potentially surprising result (Theorem~\ref{theorem:main} in Section~\ref{sec:common-knowledge}) that after the initial nondeterministic action of ``morning arrives'', regardless of how many times consecutively the human presses the ``reset" button, the human and the AI will \emph{never} attain common knowledge of the alarm being off.


\section{Background}
\label{sec:background}

In this section, we introduce the formal notions and notation needed to state and prove the unattainability of common knowledge in our scenario.

\subsection{Games with imperfect information}
\label{subsec:games}

Our focus here is on games over graphs where agents possess imperfect information, that is, the agents cannot distinguish between certain game states. This can for instance be caused by limitations of the sensors of the agents. Below we define multi-agent games with imperfect information against nature (MAGIIAN), following~\cite{gur-gor-lun-22-aij}.

\begin{definition}[MAGIIAN]
A multi-agent game with imperfect information against Nature (MAGIIAN) is a tuple $G = (Agt, Loc, l_{init}, Act, \Delta, Obs)$, where:
\begin{itemize}
    \item $\mathit{Agt} = \{a_1, a_2, \ldots, a_n\}$ is a set of agents.
    \item $Loc$ is a set of game states, called locations, usually assumed finite.
    \item For each $a_i \in Agt$, $Act_{a_i}$ is a finite set of possible actions of agent $a_i$.
    \item $Act = Act_{a_1} \times . . . \times Act_{a_n}$ are the possible action profiles (or joint actions) of the team.
    \item $\Delta \subseteq Loc \times Act \times Loc$ is a transition relation between locations, with transitions labelled by action profiles.
    \item For each $a_i \in Agt$, $Obs_{a_i}$ is a partition of~$Loc$, the blocks of which are the possible observations of agent $a_i$. Given any location $l$, the unique observation for i containing l is denoted by $obs_{a_i}(l)$. We denote with $\sim_{a_i}$ the equivalence relation on locations induced by the partition.
    \item $Obs = Obs_{a_1} \times . . . \times Obs_{a_n}$ is the set of all observation profiles (or joint observations) of the team $Agt$. 
\end{itemize}
\end{definition}

\noindent
It is assumed that all actions are available to the agents at every location.

An example, two-agent MAGIIAN is the game depicted in Figure~\ref{fig:state_transition}. 
The set of agents consists of the human and the AI. 
The actions of the human are ``morning arrives'', ``reset'', and ``toggle'', abbreviated as~$i$, $r$, and~$t$, respectively.
The AI conceptually has no actions, which is modelled by the ``dummy'' action~$\ast$. 
Therefore, as a convention, we shall identify joint actions of the team with the actions of the human. 

\begin{definition}[Full play]
    In the context of a MAGIIAN, $G$, a full play is an infinite sequence of alternating locations and actions $\pi = l_0 \sigma_1 l_1 \sigma_2 l_2 ...$, where $l_0 = l_{init}$, $\sigma_j \in Act$ and $(l_j, \sigma_{j+1}, l_{j+1}) \in \Delta$ for all $j \geq 0$.
\end{definition}

\begin{definition}[Full history]
    A full history is a finite prefix of a full play: $\pi(j) = \pi = l_0 \sigma_1 l_1 \sigma_2 l_2 ... \sigma_j l_j$. 
\end{definition}
Note that we abbreviate ``full history" to ``history", although in some works they are separate concepts.
For a history~$h$, let $ \mathit{last} (h)$ denote its last location.

\subsection{Epistemic logic}
\label{subsec:logic}

Epistemic logic allows us to reason about what agents know. In our game, the human and AI have different perspectives on the game state, and epistemic logic helps us capture how these perspectives evolve.

We now summarise the fundamentals of epistemic logic, on which this paper builds, adapted from~\cite{HuthRyanLogicInComputerScience}.
Epistemic logic is built on top of the multi-modal logic KT45$^n$.

\begin{definition}[KT45$^n$ syntax]
    The formulas of KT45$^n$ are defined by the following grammar:
    $$ \varphi ::= p \>|\> 
                   \neg \varphi \>|\> 
                   \varphi \wedge \varphi \>|\> 
                   K_i\, \varphi \>|\> 
                   E\, \varphi \>|\> 
                   C\, \varphi $$
\noindent
where $p$ ranges over a given set~$\mathit{Atoms}$ of atomic propositions.
\end{definition}

\noindent
Intuitively, $K_i\, \varphi$ expresses that agent~$a_i$ knows that~$\varphi$ is true, $E\, \varphi$ that \emph{every} agent knows that~$\varphi$ is true, and $C\, \varphi$ that the agents have \emph{common knowledge} that~$\varphi$ is true.
Note that KT45$^n$ typically includes the \emph{distributed knowledge} operator~$D\, \varphi$. We shall not need this operator for our results, and we therefore omit it from our presentation.

The formal semantics of Epistemic logic is expressed in terms of \emph{Kripke structures}. 
Let $\mathit{Agt} = \{a_1, a_2, \ldots, a_n\}$ be a set of agents.

\begin{definition}[Kripke structure]
\label{def:kripke_general}
    A \emph{Kripke structure} (or simply \emph{model}) is a triple $\mathcal{M} = (W, \{R_{a_i}\}_{a_i \in \mathit{Agt}}, L)$ consisting of:
    \begin{itemize}
        \item a set $W$ of states, called possible worlds;
        \item for each $a_i \in \mathit{Agt}$, an equivalence relation $R_{a_i}$ on $W$, called the accessibility relations;
        \item a labelling function $L: W \rightarrow \mathcal{P}(\mathit{Atoms})$.
    \end{itemize}
\end{definition}

\begin{definition}[KT45$^n$ semantics]\label{def:kt45-semantics}
    The \emph{semantics} of the propositional logic operators is standard. We only give the semantics of the epistemic operators: \\
    
    $\mathcal{M}, w \models K_i\, \varphi$ ~if~ $\mathcal{M}, v \models \varphi$ for all $v$ such that $w \sim_i v$

    ~

    $ \begin{array}{rcl}
      E\, \varphi & \equiv & 
          \bigwedge_{i \in \mathit{Agt}} K_i\, \varphi \\
      E^0\, \varphi & \equiv & 
          \varphi \\
      E^{k+1}\, \varphi & \equiv & 
          E E^k\, \varphi \\
      C\, \varphi & \equiv & 
          \bigwedge_{k=1}^{\infty} E^k\, \varphi \\
      \end{array} $
\end{definition}

We instantiate the above general notions to MAGIIAN games as follows.
We take the set of game locations to serve as the atomic propositions of the logic.

\begin{definition}[Induced model]
\label{def:induced-model}
    Let $G$ be a MAGIIAN. 
    It induces the model
    $\mathcal{M}_G = (H_G, \{\sim_{a_i}\}_{a_i \in Agt}, L)$, where:
    \begin{itemize}
        \item $H_G$ is the set of all histories of~$G$;
        \item for each agent $a_i \in \mathit{Agt}$, $\sim_{a_i}$ is an equivalence relation on histories that are \emph{indistinguishable} for the agent, i.e., histories in which it has taken the same actions and has made the same observations:
        
         $l_{0} \sigma_1 l_1 \sigma_2 ... \sigma_k l_k \>\sim_{a_i}\> l_{0}' \sigma_1' l_1' \sigma_2' ... \sigma_{k'}' l_{k'}'   
        \>\equiv\>$
    
        $k = k' \land 
        \forall j \in \{0, ..., k\}\!.\ l_j \sim_{a_i} l_j' \land \sigma_j[i] = \sigma_j'[i] $
        \item $L: H_G \rightarrow \mathcal{P} (Loc)$ is defined by: $L(h) = \{\mathit{last} (h)\}$.
    \end{itemize}
\end{definition}


\section{An iterative model update construction}

Our main theorem, in Section~\ref{sec:common-knowledge}, concerns the knowledge of our two agents, expressed as a formula of Epistemic logic, \emph{after} a given history~$h$. 
Note that the truth value of $\mathcal{M}_G, h \models \varphi$ solely depends on the \emph{strongly connected component} of~$H_G$ w.r.t.\ $\cup \{\sim_{a_i}\}_{a_i \in Agt}$ that contains~$h$.
Hence, one typically defines an \emph{epistemic update function}, which computes a new SCC based on the current one, the joint action taken, and the joint observation made upon it. 

However, in the concrete game from Section~\ref{game_set_up}, we can observe that every SCC consists of \emph{all} histories of a certain length. This allows a simplified treatment of Epistemic logic formulas in the context of this game.
First, we shall denote by~$\mathcal{M}_n$ the Kripke structure induced by the set~$H_n$ of histories having exactly $n$~actions. 
Moreover, instead of a knowledge update function, we shall present an \emph{iterative construction}, producing model~$\mathcal{M}_{n+1}$ from~$\mathcal{M}_n$, for any~$n$. 
Our \emph{model update} adds all actions to all paths to get the paths of the updated model.
The update is thus built by adding actions~$r$ and~$t$ to all paths in~$H_n$, producing $H_n^r$ and $H_n^t$, respectively, the union of which gives the set of all paths of $\mathcal{M}_{n+1}$, $H_{n+1}$.
Note that $|H_n| = 2^n$, and that the last action of a history fully determines its last location (in this particular game).

Our iterative construction will make use of the following result.

\begin{lemma}
    In our game, the equivalence relations on the histories satisfy the following inductive properties:
    $$ \begin{array}{rcl}
       \mathit{On} \sim_{Human} \mathit{Off} && \\
       h_1 \cdot a \cdot l_1 \sim_{Human} h_2 \cdot b \cdot l_2 & \Leftrightarrow & h_1 \sim_{Human} h_2 \wedge a = b \\
       h_1 \cdot a \cdot l_1 \sim_{AI} h_2 \cdot b \cdot l_2 & \Leftrightarrow &  h_1 \sim_{AI} h_2 \wedge l_1 = l_2 \\
       \end{array} $$
    where $h_1, h_2 \in H_G$ and $a, b \in Act$. 
    \label{lem:relations}
\end{lemma}

\noindent
The above properties can be seen and used as a definition of the equivalence relations by \emph{structural induction}.

Let us now illustrate our construction, starting with $\mathcal{M}_1$.
As a notational convenience, we shall omit in histories the initial location and action. 
Below are the new paths as we add the actions~$r$ and~$t$: \\

$\textit{Off} \cdot r \>\leadsto\> \text{Off r Off}$

$\textit{On} \cdot r \>\leadsto\> \text{On r Off}$

$\textit{On} \cdot t \>\leadsto\> \text{On t Off}$

$\textit{Off} \cdot t \>\leadsto\> \text{Off t On}$ \\

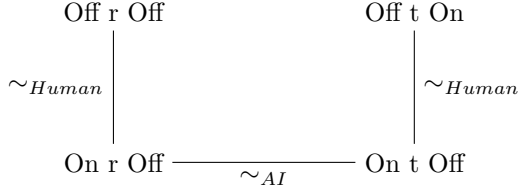
\begin{figure}[t]
    \centering
    \begin{tikzpicture}
      \node (A) at (0,0) {Off r Off};
      \node (B) at (4,0) {Off t On};
      \node (C) at (0,-2) {On r Off};
      \node (D) at (4,-2) {On t Off};
    
      \draw (A) -- node[left] {$\sim_{Human}$} (C);
      \draw (B) -- node[right] {$\sim_{Human}$} (D);
      \draw (C) -- node[below] {$\sim_{AI}$} (D);
    \end{tikzpicture}
    \caption{the resulting sequences and the relations between them.}
    \label{fig:ending_relations}
\end{figure}

\noindent
Next, we want to find the new equivalence relations, using Lemma~\ref{lem:relations}. 
Exploring the relations in $H_n^r$ and $H_n^t$ separately, we can conclude that the relations are preserved, since the same nodes are appended only if the previous nodes were the same, preserving $\sim_{AI}$, and the added action are identical thus preserving $\sim_{Human}$. 
The new relations between the paths in~$H_n^r$ and~$H_n^t$ are as follows (see Figure~\ref{fig:ending_relations}).
The only relations between actions~$r$ and~$t$ are at paths that previously ended with \textit{On}. Thus all paths that were on $On$ the move before have a $\sim_{AI}$ relation between one path in $H_n^r$ and the another one  in $H_n^t$. We just need to add these relations to form $R_{n+1}$. Note that this also implies that the path $ir^n$ is only related to another node and is therefore on one end of the relation chain.

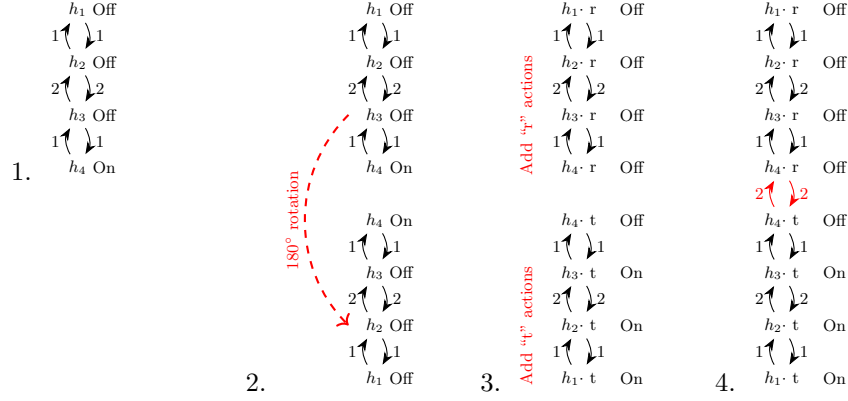
\begin{figure}[t]
    \centering
    \begin{multicols}{4}
        \begin{enumerate}
            \item \begin{tikzpicture}[scale=0.7, transform shape]
                \tikzset{every node/.style={font=\small}}
                
                \node (h1) at (0,0) {$h_1$};
                \node (text1) at (0.5,0) {Off};
                \node (h2) at (0,-1) {$h_2$};
                \node (text2) at (0.5,-1) {Off};
                \node (h3) at (0,-2) {$h_3$};
                \node (text3) at (0.5,-2) {Off};
                \node (h4) at (0,-3) {$h_4$};
                \node (text4) at (0.5,-3) {On};
                
                \draw[-{Stealth[scale=1]}] (h1) to [bend left] node[right] {1} (h2);
                \draw[-{Stealth[scale=1]}] (h2) to [bend left] node[left] {1} (h1);
                \draw[-{Stealth[scale=1]}] (h2) to [bend left] node[right] {2} (h3);
                \draw[-{Stealth[scale=1]}] (h3) to [bend left] node[left] {2} (h2);
                \draw[-{Stealth[scale=1]}] (h3) to [bend left] node[right] {1} (h4);
                \draw[-{Stealth[scale=1]}] (h4) to [bend left] node[left] {1} (h3);
            \end{tikzpicture}
            \item \begin{tikzpicture}[scale=0.7, transform shape]
                \tikzset{every node/.style={font=\small}}
                
                \node (h1) at (0,0) {$h_1$};
                \node (text1) at (0.5,0) {Off};
                \node (h2) at (0,-1) {$h_2$};
                \node (text2) at (0.5,-1) {Off};
                \node (h3) at (0,-2) {$h_3$};
                \node (text3) at (0.5,-2) {Off};
                \node (h4) at (0,-3) {$h_4$};
                \node (text4) at (0.5,-3) {On};
                \node (h5) at (0,-7) {$h_1$};
                \node (text1) at (0.5,-7) {Off};
                \node (h6) at (0,-6) {$h_2$};
                \node (text2) at (0.5,-6) {Off};
                \node (h7) at (0,-5) {$h_3$};
                \node (text3) at (0.5,-5) {Off};
                \node (h8) at (0,-4) {$h_4$};
                \node (text4) at (0.5,-4) {On};
                
                \draw[-{Stealth[scale=1]}] (h1) to [bend left] node[right] {1} (h2);
                \draw[-{Stealth[scale=1]}] (h2) to [bend left] node[left] {1} (h1);
                \draw[-{Stealth[scale=1]}] (h2) to [bend left] node[right] {2} (h3);
                \draw[-{Stealth[scale=1]}] (h3) to [bend left] node[left] {2} (h2);
                \draw[-{Stealth[scale=1]}] (h3) to [bend left] node[right] {1} (h4);
                \draw[-{Stealth[scale=1]}] (h4) to [bend left] node[left] {1} (h3);
                \draw[-{Stealth[scale=1]}] (h5) to [bend left] node[left] {1} (h6);
                \draw[-{Stealth[scale=1]}] (h6) to [bend left] node[right] {1} (h5);
                \draw[-{Stealth[scale=1]}] (h6) to [bend left] node[left] {2} (h7);
                \draw[-{Stealth[scale=1]}] (h7) to [bend left] node[right] {2} (h6);
                \draw[-{Stealth[scale=1]}] (h7) to [bend left] node[left] {1} (h8);
                \draw[-{Stealth[scale=1]}] (h8) to [bend left] node[right] {1} (h7);
    
                \draw[->, thick, dashed, red] (-0.5,-2) to [bend right=45] node[midway, above, rotate=90] {180$^\circ$ rotation} (-0.5,-6);
            \end{tikzpicture}
            \item \begin{tikzpicture}[scale=0.7, transform shape]
                \tikzset{every node/.style={font=\small}}
                
                \node (h1) at (0,0) {$h_1 \cdot$ r};
                \node (text1) at (1,0) {Off};
                \node (h2) at (0,-1) {$h_2 \cdot$ r};
                \node (text2) at (1,-1) {Off};
                \node (h3) at (0,-2) {$h_3 \cdot$ r};
                \node (text3) at (1,-2) {Off};
                \node (h4) at (0,-3) {$h_4 \cdot$ r};
                \node (text4) at (1,-3) {Off};
                \node (h5) at (0,-7) {$h_1 \cdot$ t};
                \node (text1) at (1,-7) {On};
                \node (h6) at (0,-6) {$h_2 \cdot$ t};
                \node (text2) at (1,-6) {On};
                \node (h7) at (0,-5) {$h_3 \cdot$ t};
                \node (text3) at (1,-5) {On};
                \node (h8) at (0,-4) {$h_4 \cdot$ t};
                \node (text4) at (1,-4) {Off};
                
                \draw[-{Stealth[scale=1]}] (h1) to [bend left] node[right] {1} (h2);
                \draw[-{Stealth[scale=1]}] (h2) to [bend left] node[left] {1} (h1);
                \draw[-{Stealth[scale=1]}] (h2) to [bend left] node[right] {2} (h3);
                \draw[-{Stealth[scale=1]}] (h3) to [bend left] node[left] {2} (h2);
                \draw[-{Stealth[scale=1]}] (h3) to [bend left] node[right] {1} (h4);
                \draw[-{Stealth[scale=1]}] (h4) to [bend left] node[left] {1} (h3);
                \draw[-{Stealth[scale=1]}] (h5) to [bend left] node[left] {1} (h6);
                \draw[-{Stealth[scale=1]}] (h6) to [bend left] node[right] {1} (h5);
                \draw[-{Stealth[scale=1]}] (h6) to [bend left] node[left] {2} (h7);
                \draw[-{Stealth[scale=1]}] (h7) to [bend left] node[right] {2} (h6);
                \draw[-{Stealth[scale=1]}] (h7) to [bend left] node[left] {1} (h8);
                \draw[-{Stealth[scale=1]}] (h8) to [bend left] node[right] {1} (h7);
    
                \node[rotate=90, red] at (-1,-2) {Add ``r'' actions};
                \node[rotate=90, red] at (-1,-6) {Add ``t'' actions};
            \end{tikzpicture}
            \item \begin{tikzpicture}[scale=0.7, transform shape]
                \tikzset{every node/.style={font=\small}}
                
                \node (h1) at (0,0) {$h_1 \cdot$ r};
                \node (text1) at (1,0) {Off};
                \node (h2) at (0,-1) {$h_2 \cdot$ r};
                \node (text2) at (1,-1) {Off};
                \node (h3) at (0,-2) {$h_3 \cdot$ r};
                \node (text3) at (1,-2) {Off};
                \node (h4) at (0,-3) {$h_4 \cdot$ r};
                \node (text4) at (1,-3) {Off};
                \node (h5) at (0,-7) {$h_1 \cdot$ t};
                \node (text1) at (1,-7) {On};
                \node (h6) at (0,-6) {$h_2 \cdot$ t};
                \node (text2) at (1,-6) {On};
                \node (h7) at (0,-5) {$h_3 \cdot$ t};
                \node (text3) at (1,-5) {On};
                \node (h8) at (0,-4) {$h_4 \cdot$ t};
                \node (text4) at (1,-4) {Off};
                
                \draw[-{Stealth[scale=1]}] (h1) to [bend left] node[right] {1} (h2);
                \draw[-{Stealth[scale=1]}] (h2) to [bend left] node[left] {1} (h1);
                \draw[-{Stealth[scale=1]}] (h2) to [bend left] node[right] {2} (h3);
                \draw[-{Stealth[scale=1]}] (h3) to [bend left] node[left] {2} (h2);
                \draw[-{Stealth[scale=1]}] (h3) to [bend left] node[right] {1} (h4);
                \draw[-{Stealth[scale=1]}] (h4) to [bend left] node[left] {1} (h3);
                \draw[-{Stealth[scale=1]}] (h5) to [bend left] node[left] {1} (h6);
                \draw[-{Stealth[scale=1]}] (h6) to [bend left] node[right] {1} (h5);
                \draw[-{Stealth[scale=1]}] (h6) to [bend left] node[left] {2} (h7);
                \draw[-{Stealth[scale=1]}] (h7) to [bend left] node[right] {2} (h6);
                \draw[-{Stealth[scale=1]}] (h7) to [bend left] node[left] {1} (h8);
                \draw[-{Stealth[scale=1]}] (h8) to [bend left] node[right] {1} (h7);
    
                \draw[-{Stealth[scale=1]}, red] (h8) to [bend left] node[left] {2} (h4);
                \draw[-{Stealth[scale=1]}, red] (h4) to [bend left] node[right] {2} (h8);
            \end{tikzpicture}
        \end{enumerate}
    \end{multicols}
    \caption{Model update, applied to~$\mathcal{M}_2$ (step 1), resulting in~$\mathcal{M}_3$ (step 4).}
    \label{fig:update_function_steps}
\end{figure}

This makes it natural to see model update as the following \emph{informal procedure}, illustrated in Figure~\ref{fig:update_function_steps}:

\begin{enumerate}
    \item Take the paths in $H_n$ and keep them ordered by the relation correspondence switching between $\sim_{Human}$ and $\sim_{AI}$, making sure that all paths are in a row. One will see that all paths ending with \textit{On} and \textit{Off} will be in a row.
    \item Start at the end where the \textit{On}'s are and add all paths by "rotation" at 180 degrees.
    \item Add $r$'s to the upper half and $t$'s to the lower half.
    \item Draw all old relations and draw new $\sim_{AI}$ relations by starting in the middle where all the \textit{On}'s were (which now are \textit{Off}'s). Connect the paths one step to the right and left of the centre, then continue with the second, third, and so on.
\end{enumerate}

Finally, we can formally characterise the iterative construction.

\begin{lemma}[Model update]
\label{lem:update}
Given model $\mathcal{M}_{n} = (H_n, \{\sim_{a_i}\}_{a_i \in Agt}, L)$, the updated model is the model $\mathcal{M}_{n+1} = (H_{n+1}, \{\sim_{a_i}'\}_{a_i \in Agt}, L)$, where the new equivalence relations are defined as follows.

For histories \( h, h' \in H_n \), \( h\cdot r \sim_{a_i}
' h'^r \) if and only if \( h \sim_{a_i} h' \), and similarly, \( h\cdot t \sim_{a_i}' h'\cdot t \) if and only if \( h \sim_{a_i} h' \). This ensures that the relations within each action-specific set are preserved exactly as they were in the original set~$H_n$. For the AI agent, the cross-relation \( \sim_{AI}' \) between histories in $H_n^r$ and $H_n^t$ is defined based on the continuity of game states concerning the AI's perception capabilities:
    \[
    h\cdot r \sim_{AI}' h\cdot t \quad \text{if and only if} \quad L(h) = \text{"On"}
    \]

\end{lemma}

\begin{figure}[t]
    \centering
    \begin{tikzpicture}[scale=1, transform shape]
        \tikzset{every node/.style={font=\small}}
        
        \node (h1) at (0,0) {$h_1$};
        \node (text1) at (1,0) {Off};
        \node (h2) at (0,-1) {$h_2$};
        \node (text2) at (1,-1) {Off};
        \node (h3) at (0,-2) {$h_3$};
        \node (text3) at (1,-2) {Off};
        \node (h4) at (0,-3) {$h_4$};
        \node (text4) at (1,-3) {Off};
        \node (h8) at (0,-7) {$h_8$};
        \node (text1) at (1,-7) {On};
        \node (h7) at (0,-6) {$h_7$};
        \node (text2) at (1,-6) {On};
        \node (h6) at (0,-5) {$h_6$};
        \node (text3) at (1,-5) {On};
        \node (h5) at (0,-4) {$h_5$};
        \node (text4) at (1,-4) {Off};
        
        \draw[-{Stealth[scale=1]}] (h1) to [bend left] node[right] {1} (h2);
        \draw[-{Stealth[scale=1]}] (h2) to [bend left] node[left] {1} (h1);
        \draw[-{Stealth[scale=1]}] (h2) to [bend left] node[right] {2} (h3);
        \draw[-{Stealth[scale=1]}] (h3) to [bend left] node[left] {2} (h2);
        \draw[-{Stealth[scale=1]}] (h3) to [bend left] node[right] {1} (h4);
        \draw[-{Stealth[scale=1]}] (h4) to [bend left] node[left] {1} (h3);
        \draw[-{Stealth[scale=1]}] (h5) to [bend left] node[right] {1} (h6);
        \draw[-{Stealth[scale=1]}] (h6) to [bend left] node[left] {1} (h5);
        \draw[-{Stealth[scale=1]}] (h6) to [bend left] node[right] {2} (h7);
        \draw[-{Stealth[scale=1]}] (h7) to [bend left] node[left] {2} (h6);
        \draw[-{Stealth[scale=1]}] (h7) to [bend left] node[right] {1} (h8);
        \draw[-{Stealth[scale=1]}] (h8) to [bend left] node[left] {1} (h7);
        \draw[-{Stealth[scale=1]}, red] (h5) to [bend left] node[left] {2} (h4);
        \draw[-{Stealth[scale=1]}, red] (h4) to [bend left] node[right] {2} (h5);
    
        \begin{scope}[scale=0.45]
            \node (h1_) at (7,0) {$h_1$};
            \node (text1_) at (8,0) {Off};
            \node (h2_) at (7,-1) {$h_2$};
            \node (text2_) at (8,-1) {Off};
            \node (h3_) at (7,-2) {$h_3$};
            \node (text3_) at (8,-2) {Off};
            \node (h4_) at (7,-3) {$h_4$};
            \node (text4_) at (8,-3) {Off};
            \node (h8_) at (7,-7) {$h_8$};
            \node (text5_) at (8,-7) {Off};
            \node (h7_) at (7,-6) {$h_7$};
            \node (text6_) at (8,-6) {Off};
            \node (h6_) at (7,-5) {$h_6$};
            \node (text7_) at (8,-5) {Off};
            \node (h5_) at (7,-4) {$h_5$};
            \node (text8_) at (8,-4) {Off};
            \node (h9_) at (7,-8) {$h_9$};
            \node (text9_) at (8,-8) {Off};
            \node (h10_) at (7,-9) {$h_{10}$};
            \node (text10_) at (8,-9) {Off};
            \node (h11_) at (7,-10) {$h_{11}$};
            \node (text11_) at (8,-10) {Off};
            \node (h12_) at (7,-11) {$h_{12}$};
            \node (text12_) at (8,-11) {On};
            \node (h13_) at (7,-12) {$h_{13}$};
            \node (text13_) at (8,-12) {On};
            \node (h14_) at (7,-13) {$h_{14}$};
            \node (text14_) at (8,-13) {On};
            \node (h15_) at (7,-14) {$h_{15}$};
            \node (text15_) at (8,-14) {On};
            \node (h16_) at (7,-15) {$h_{16}$};
            \node (text16_) at (8,-15) {On};
            
            \draw[-{Stealth[scale=0.5]}] (h1_) to [bend left] node[right] {1} (h2_);
            \draw[-{Stealth[scale=0.5]}] (h2_) to [bend left] node[left] {1} (h1_);
            \draw[-{Stealth[scale=0.5]}] (h2_) to [bend left] node[right] {2} (h3_);
            \draw[-{Stealth[scale=0.5]}] (h3_) to [bend left] node[left] {2} (h2_);
            \draw[-{Stealth[scale=0.5]}] (h3_) to [bend left] node[right] {1} (h4_);
            \draw[-{Stealth[scale=0.5]}] (h4_) to [bend left] node[left] {1} (h3_);
            \draw[-{Stealth[scale=0.5]}] (h5_) to [bend left] node[right] {1} (h6_);
            \draw[-{Stealth[scale=0.5]}] (h6_) to [bend left] node[left] {1} (h5_);
            \draw[-{Stealth[scale=0.5]}] (h6_) to [bend left] node[right] {2} (h7_);
            \draw[-{Stealth[scale=0.5]}] (h7_) to [bend left] node[left] {2} (h6_);
            \draw[-{Stealth[scale=0.5]}] (h7_) to [bend left] node[right] {1} (h8_);
            \draw[-{Stealth[scale=0.5]}] (h8_) to [bend left] node[left] {1} (h7_);
            \draw[-{Stealth[scale=0.5]}] (h5_) to [bend left] node[left] {2} (h4_);
            \draw[-{Stealth[scale=0.5]}] (h4_) to [bend left] node[right] {2} (h5_);
            \draw[-{Stealth[scale=0.5]}] (h9_) to [bend left] node[right] {1} (h10_);
            \draw[-{Stealth[scale=0.5]}] (h10_) to [bend left] node[left] {1} (h9_);
            \draw[-{Stealth[scale=0.5]}] (h10_) to [bend left] node[right] {2} (h11_);
            \draw[-{Stealth[scale=0.5]}] (h11_) to [bend left] node[left] {2} (h10_);
            \draw[-{Stealth[scale=0.5]}] (h11_) to [bend left] node[right] {1} (h12_);
            \draw[-{Stealth[scale=0.5]}] (h12_) to [bend left] node[left] {1} (h11_);
            \draw[-{Stealth[scale=0.5]}] (h13_) to [bend left] node[right] {1} (h14_);
            \draw[-{Stealth[scale=0.5]}] (h14_) to [bend left] node[left] {1} (h13_);
            \draw[-{Stealth[scale=0.5]}] (h14_) to [bend left] node[right] {2} (h15_);
            \draw[-{Stealth[scale=0.5]}] (h15_) to [bend left] node[left] {2} (h14_);
            \draw[-{Stealth[scale=0.5]}] (h15_) to [bend left] node[right] {1} (h16_);
            \draw[-{Stealth[scale=0.5]}] (h16_) to [bend left] node[left] {1} (h15_);
            \draw[-{Stealth[scale=0.5]}] (h13_) to [bend left] node[left] {2} (h12_);
            \draw[-{Stealth[scale=0.5]}] (h12_) to [bend left] node[right] {2} (h13_);
    
            \draw[-{Stealth[scale=0.5]}, red] (h9_) to [bend left=45] node[left] {2} (h8_);
            \draw[-{Stealth[scale=0.5]}, red] (h8_) to [bend right=45] node[left] {2} (h9_);
            \draw[-{Stealth[scale=0.5]}, red] (h10_) to [bend left=45] node[left] {2} (h7_);
            \draw[-{Stealth[scale=0.5]}, red] (h7_) to [bend right=45] node[left] {2} (h10_);
            \draw[-{Stealth[scale=0.5]}, red] (h11_) to [bend left=45] node[left] {2} (h6_);
            \draw[-{Stealth[scale=0.5]}, red] (h6_) to [bend right=45] node[left] {2} (h11_);
        \end{scope}
    
        \begin{scope}[scale=0.23]
            \foreach \i in {1,...,32} {
                \node (h\i__) at (26,{-1*(\i-1)}) {$h_{\i}$};
            }
            \foreach \i in {1,...,21} {
                \node (text\i__) at (27,{-1*(\i-1)}) {Off};
            }
            \foreach \i in {22,...,32} {
                \node (text\i__) at (27,{-1*(\i-1)}) {On};
            }
    
            \foreach \i/\j in {1/2, 3/4, 5/6, 7/8, 9/10, 11/12, 13/14, 15/16, 17/18, 19/20, 21/22, 23/24, 25/26, 27/28, 29/30, 31/32} {
                \draw[-{Stealth[scale=0.25]}] (h\i__) to [bend left] node[right] {1} (h\j__);
                \draw[-{Stealth[scale=0.25]}] (h\j__) to [bend left] node[left] {1} (h\i__);
            }
    
            \foreach \i/\j in {2/3, 4/5, 6/7, 8/9, 10/11, 12/13, 14/15, 18/19, 20/21, 22/23, 24/25, 26/27, 28/29, 30/31} {
                \draw[-{Stealth[scale=0.25]}] (h\i__) to [bend left] node[right] {2} (h\j__);
                \draw[-{Stealth[scale=0.25]}] (h\j__) to [bend left] node[left] {2} (h\i__);
            }
    
            \draw[-{Stealth[scale=0.5]}] (h10__) to [bend left=45] node[left] {2} (h7__);
            \draw[-{Stealth[scale=0.5]}] (h7__) to [bend right=45] node[left] {2} (h10__);
            \draw[-{Stealth[scale=0.5]}] (h11__) to [bend left=45] node[left] {2} (h6__);
            \draw[-{Stealth[scale=0.5]}] (h6__) to [bend right=45] node[left] {2} (h11__);
    
            \draw[-{Stealth[scale=0.5]}] (h23__) to [bend right=45] node[right] {2} (h26__);
            \draw[-{Stealth[scale=0.5]}] (h26__) to [bend left=45] node[right] {2} (h23__);
            \draw[-{Stealth[scale=0.5]}] (h22__) to [bend right=45] node[right] {2} (h27__);
            \draw[-{Stealth[scale=0.5]}] (h27__) to [bend left=45] node[right] {2} (h22__);
    
            \foreach \i/\j in {17/16, 18/15, 19/14, 20/13, 21/12} {
                \draw[-{Stealth[scale=0.5]}, red] (h\i__) to [bend left=45] node[left] {2} (h\j__);
            }
        \end{scope}
    \end{tikzpicture}
    \caption{ $\mathcal{M}_{1}$ with $h_1=\text{Start i Off r Off}$ \\
        \hspace*{11mm} $\leadsto \mathcal{M}_{2}$ with $h_1=\text{Start i Off r Off r Off}$ \\
        \hspace*{11mm} $\leadsto \mathcal{M}_{3}$ with $h_1=\text{Start i Off r Off r Off r Off}$
    }
    \label{fig:update_visualisation}
\end{figure}
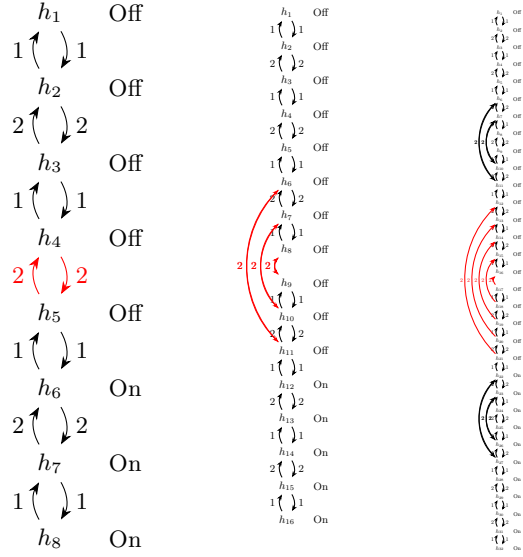

It is interesting to note that jumps appear due to the new $\sim_{AI}$ relations, and the jumps are preserved in the later updates. This causes the emergence of more and more jumps as can be seen in Figure~\ref{fig:update_visualisation}.

\section{Unattainability of common knowledge}
\label{sec:common-knowledge}

To answer the question, posed in Section~\ref{game_set_up}, of whether the AI can know the knowledge state of the human, we prove that common knowledge is unattainable for certain action sequences. Thus, in certain cases, the AI cannot know the human's knowledge state. 

We now aim to prove this statement. Namely, that for certain histories~$h \in H_n$, the condition:
$$ \mathcal{M}_n, h \models C\, \mathit{Off} $$ 
does \emph{not} hold. Using our model update Lemma~\ref{lem:update}, which is essential for the proof, this can be shown rather easily.

\begin{figure}[t]
    \centering
    \begin{tikzpicture}[scale=0.8, transform shape]
        \tikzset{every node/.style={font=\small}}
    
        \node (h1) at (0,0) {$h_1$};
        \node (text1) at (1,0) {Off};
        \node (vdots1) at (0,-0.5) {$\vdots$};
        \node (h2) at (0,-1) {$h_i$};
        \node (text2) at (1,-1) {Off};
        \node (h3) at (0,-1.5) {$h_{i+1}$};
        \node (text3) at (1,-1.5) {On};
        \node (vdots2) at (0,-2) {$\vdots$};
        \node (h4) at (0,-2.5) {$h_{j}$};
        \node (text4) at (1,-2.5) {On};
    
        \node (h8) at (0,-3.5) {$h_{j}$};
        \node (text5) at (1,-3.5) {On};
        \node (vdots3) at (0,-4) {$\vdots$};
        \node (h7) at (0,-4.5) {$h_{i+1}$};
        \node (text6) at (1,-4.5) {On};
        \node (h6) at (0,-5) {$h_i$};
        \node (text7) at (1,-5) {Off};
        \node (vdots4) at (0,-5.5) {$\vdots$};
        \node (h5) at (0,-6) {$h_1$};
        \node (text8) at (1,-6) {Off};
    
        \node (text_1) at (2.5,0) {Off};
        \node (text_2) at (2.5,-1) {Off};
        \node (text_3) at (2.5,-1.5) {Off};
        \node (text_4) at (2.5,-2.5) {Off};
        \node (text_5) at (2.5,-3.5) {Off};
        \node (text_6) at (2.5,-4.5) {Off};
        \node (text_7) at (2.5,-5) {On};
        \node (text_8) at (2.5,-6) {On};
    
        \node (text_) at (1.7,0.5) {Update};
        \node (text_) at (3.3,0.5) {Jumps};
    
        \draw[-{Stealth[scale=0.5]}] (text1) to node[right] {} (text_1);
        \draw[-{Stealth[scale=0.5]}] (text2) to node[right] {} (text_2);
        \draw[-{Stealth[scale=0.5]}] (text3) to node[right] {} (text_3);
        \draw[-{Stealth[scale=0.5]}] (text4) to node[right] {} (text_4);
        \draw[-{Stealth[scale=0.5]}] (text5) to node[right] {} (text_5);
        \draw[-{Stealth[scale=0.5]}] (text6) to node[right] {} (text_6);
        \draw[-{Stealth[scale=0.5]}] (text7) to node[right] {} (text_7);
        \draw[-{Stealth[scale=0.5]}] (text8) to node[right] {} (text_8);
    
        \draw[-{Stealth[scale=0.5]}] (text_1) to [bend left=85] node[right] {2n+1} (text_3);
        \draw[-{Stealth[scale=0.5]}] (text_3) to [bend left=85] node[right] {1; large jump} (text_6);
        \draw[-{Stealth[scale=0.5]}] (text_6) to [bend left=85] node[right] {1} (text_7);
    
    \end{tikzpicture}
    \caption{Visualisation of why only two jumps are required after the update.}
    \label{fig:proof_end_visualisation}
\end{figure}
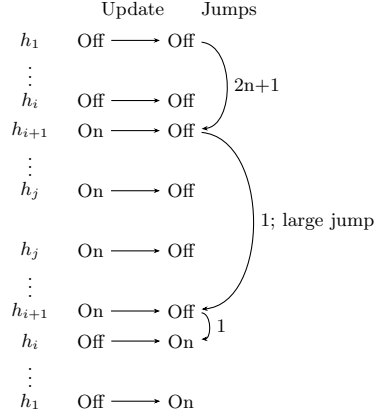

\begin{theorem}\label{theorem:main}
For every $n \in \mathbb{N}$ and history~$h$ over action sequence~$i r^n$, we have:  
$$ \mathcal{M}_n, h \models \forall k \leq n.\, 
   (E^{2k} \mathit{Off} \wedge \neg E^{2n + 1} \mathit{Off}) $$
\end{theorem}

\begin{proof}
By mathematical induction on $n$.\\

\noindent
\textbf{Base case.} 
Let $n=0$. 
Then $2n+1 = 1$, and we therefore only have to consider the case when $k = 0$. 
In this case $\mathcal{M}_n$ consists of the two histories $h_1 = \text{Start i On}$ and $h_2 = \text{Start i Off}$, which are indistinguishable for the human, but distinguishable for the AI. 
For both histories, $E^0 \mathit{Off}$ holds, since by definition \ref{def:kt45-semantics} $E^0 \mathit{Off} \equiv \mathit{Off}$ and $\mathit{Off}$ is true due to both ending with location~$\mathit{Off}$. 
Furthermore, $\neg E^1 \mathit{Off}$ holds because $K_\mathit{Human} \mathit{Off}$ does not hold, as the human cannot distinguish whether the alarm is off or on. This establishes the base case. \\

\noindent
\textbf{Induction.} 
Assume that the result holds for~$n$ (induction hypothesis). Specifically, assume that in $\mathcal{M}_n$, the shortest path (in the chain of histories) from the first \textit{Off} state, corresponding to $i r^n$, to the first \textit{On} state is $2n + 1$.
When updating $\mathcal{M}_{n}$ to $\mathcal{M}_{n+1}$, all \textit{On}'s (which lie in the middle of equivalence chains due to the rotation placing all \textit{On}'s in the centre) become \textit{Off}'s. The histories corresponding to the actions $i r^n$ are positioned at the edges of the equivalence chain in $\mathcal{M}_{n+1}$, with $r$ or $t$ added, while only $i r^{n+1}$ retains the state \textit{Off}. The new \textit{Off} states created during the model update act as an additional ``obstacle'' between the history $i r^n$ and the shortest path to the first history with \textit{On} as the last state in $\mathcal{M}_n$. However, as illustrated in Figure~\ref{fig:proof_end_visualisation}, that whole intermediate part can be bypassed, by just using the $\sim_{AI}$ relations that are created by the model update according to Lemma~\ref{lem:update}. The number of such jumps required is exactly two, corresponding to the two new equivalence relations introduced by the update process. Thus, we just need to take two more jumps in $\mathcal{M}_{n+1}$ than in~$\mathcal{M}_{n}$. Hence, the shortest path in $\mathcal{M}_{n+1}$ is $2(n+1)+1$, which is what we needed to prove for the induction case.
\end{proof}

Our proof shows that for all histories resulting from action sequences of the form $ir^n$, \emph{common knowledge of the current location is unattainable}. For our game, this means that the AI and the human cannot reach the same knowledge state, even if the human presses the ``reset'' button consecutively an unbounded number of times. This answers the previously asked question, ``Can the AI always truly know the knowledge state of the observant human?'', with a resounding ``No''. 

Even more, by Theorem~\ref{theorem:main}, if the AI is limited in that it can reason at epistemic nesting depths of, say, at most 10, then with just 5 consecutive ``reset'' button presses the human can ``trick'' the AI into an incorrect knowledge state.


\section{Related Work}

The problem we consider in this paper appears to be related to the well-known \emph{Coordinated Attack Problem}~\cite{FaginReasoningAboutKnowledge}, which poses the question of whether common knowledge, a prerequisite for action coordination, is attainable in games with unreliable communication (modelled with a ``lossy" channel) between agents. As in our case, it turns out that it is not. Conceptually speaking, in our game there also is a ``lossy'' channel of a sort, if we view the actions of the human as an indirect channel of communication to the AI. Then, information is lost when the game switches from state \textit{On} to state \textit{Off}, since the AI cannot know whether the action taken was ``reset'' or ``toggle''. However, in our case the loss of information is due to the fundamental asymmetry in the abilities of the two agents, which is not the case in the Coordinated Attack Problem.

Recent studies, such as Bergmark et al.~\cite{Bergmark2022InvestigationStability}, have examined multi-agent systems under limited communication conditions similar to ours, investigating the stability and robustness of such systems. While their focus has been on stability and decision-making efficiency in imperfect information settings, they do not explore the unattainability of common knowledge to the extent that this paper does. We study the extreme case where no communication is allowed, further emphasising the role of asymmetry in the failure to achieve mutual understanding.

Yanjun Li’s work on multi-agent conformant planning with group knowledge proposes a dynamic epistemic framework to capture the evolution of knowledge in systems where agents are unable to observe~\cite{li2023complexity}. This study formalises multi-agent conformant planning as a model-checking problem, demonstrating that the problem is PSPACE-complete relative to the size of dynamic epistemic models. Li’s approach bridges the gap between epistemic planning and formal verification by showing that for every Kripke model with perfect recall and no miracles, there exists an equivalent dynamic epistemic model. The focus on reducing planning problems to model-checking complexities distinguishes this work from others that focus more on the strategic dynamics of knowledge asymmetry rather than on computational complexity.

Liu’s contribution on multi-agent epistemic planning with common knowledge~\cite{liu2018epistemic} presents a different but complementary perspective. Their work builds on the dynamics of belief revision in multi-agent systems, offering a framework for reasoning about higher-order beliefs and knowledge changes. Similar to our work, Liu investigates the difficulty of achieving common knowledge, but the focus is on providing a computational model for belief updates and decision-making processes in planning domains. Both studies emphasise the limitations posed by epistemic constraints, but our focus extends to the extreme case where common knowledge is fundamentally unattainable due to inherent asymmetries in information access.

Pietarinen’s work on games and logics of knowledge~\cite{pietarinen2002games} explores how agents reason about their own knowledge and the knowledge of others in strategic contexts. This foundational study, like our own, relies on Kripke structures and epistemic logic to model the evolution of knowledge states in multi-agent interactions. Pietarinen's work, however, approaches the problem from a broader theoretical lens, focusing on the logical relationships between agents' knowledge rather than the specific unattainability of common knowledge in asymmetrical games. Both works contribute to the growing literature on epistemic reasoning but differ in their treatment of asymmetry and practical game dynamics.

These studies, including foundational works by Fagin et al.~\cite{FaginReasoningAboutKnowledge} and Rubinstein~\cite{Rubinstein_1989}, collectively form the theoretical basis of our investigation, providing diverse perspectives on how knowledge is formed, updated, and constrained in multi-agent systems.

Finally, our work contrasts with studies like those of Nylén et al.~\cite{Nyln2018InvestigationOA}, where multi-agent coordination is achieved under conditions of uncertainty but with less stringent restrictions on agent communication and observation capabilities. In this paper, we deliberately impose more severe limitations to examine the breakdown of knowledge symmetry, providing new theoretical contributions to the study of multi-agent systems with incomplete information.


\section{Conclusion}

This paper rigorously examined the dynamics of common knowledge with an asymmetric game characterised by imperfect information. The core contribution of this work is the formal proof that under certain action sequences within the described model game, common knowledge of the alarm's state remains fundamentally unattainable. This insight sets a limit on the knowledge that can be achieved in such settings, highlighting significant implications for theoretical and practical applications in game theory and strategic decision-making.

The implications of these findings are manifold. They underscore the challenges and limitations inherent in designing systems where agents must cooperate despite asymmetric capabilities. Example areas range from automated trading systems to cooperative robotics and distributed systems, where understanding the bounds of shared knowledge can influence system architecture and interaction protocols.
Our findings suggest important limitations in systems requiring agent cooperation with incomplete information, and future research could explore how small modifications in communication channels could mitigate these limitations and when these limitations arise for a more general case.


\bibliographystyle{alpha}
\bibliography{refs}

\end{document}